\documentclass[aps,prd,10pt,showpacs,showkeys,twocolumn,superscriptaddress,groupedaddress]{revtex4-1}
\usepackage{epsfig}
\usepackage{graphicx}
\usepackage{graphics}
\usepackage{xspace}
\usepackage{amssymb}
\usepackage{amsmath}
\usepackage[dvips]{color}
\usepackage{latexsym}
\usepackage{mathrsfs}
\usepackage{url}
\usepackage[english]{babel}
\usepackage[latin1]{inputenc}
\usepackage[T1]{fontenc}
\newcommand{\inieq}{\begin{eqnarray}}            
\newcommand{\fineq}{\end{eqnarray}}            
\newcommand{\be}{\begin{equation}}
\newcommand{\ee}{\end{equation}}
\newcommand{\ba}{\begin{eqnarray}}
\newcommand{\ea}{\end{eqnarray}}

\def\ee{\mbox{$\left(e,e^{\prime}\right)$\ }}
\def\eep{\mbox{$\left(e,e^{\prime}p\right)$\ }}

\begin{document}
\title{The Relativistic Green's function model and charged-current inclusive neutrino-nucleus 
scattering at T2K kinematics }
\author{Andrea Meucci} 
\author{Carlotta Giusti}
\affiliation{Dipartimento di Fisica, 
Universit\`{a} degli Studi di Pavia and \\
INFN,
Sezione di Pavia, via A. Bassi 6, I-27100 Pavia, Italy}

\date{\today}

\begin{abstract}
We compare the results of the relativistic Green's function model with the experimental data of the charged-current inclusive differential neutrino-nucleus cross sections
published by the T2K Collaboration. The model, which is able to describe both MINER$\nu$A and MiniBooNE charged-current quasielastic scattering data, 
underpredicts the inclusive T2K cross sections.
\end{abstract}


\maketitle

\section{Introduction}
\label{intro}

In recent years many Collaborations have presented very interesting
results of neutrino oscillations that aim at a precise determination of mass-squared splitting and 
mixing angles in $\nu_{\mu}$  disappearance and $\nu_{e}$ appearance measurements. 
Since modern experiments, such as  MiniBooNE, SciBooNE, ArgoNeuT, MINER$\nu$A, and T2K,
are performed with detectors made of medium-heavy nuclear targets, e.g, Carbon, Oxygen, Argon, or Iron,
a clear understanding of neutrino-nucleus interactions, where all nuclear effects are well under control, is required 
for a proper analysis of experimental data. 
 The recent progress, the questions and challenges
in the physics of neutrino cross sections are  reviewed 
in \cite{RevModPhys.84.1307,Morfin:2012kn,nieves:2014}.

The first measurements of the charged-current quasielastic (CCQE)
flux-averaged double-differential $\nu_{\mu} (\bar{\nu}_{\mu})$ cross
section on $^{12}$C in the few GeV region by the MiniBooNE 
Collaboration  \cite{miniboone,miniboone-ant} have raised  extensive
discussions that effects beyond the impulse approximation (IA) may play a 
significant role in  this energy domain  \cite{PhysRevC.79.034601,Martini:2009uj,
Martini:2010ex,Martini:2013sha,Leitner:2010kp,
PhysRevC.83.054616,FernandezMartinez2011477,AmaroAntSusa,Amaro:2011qb,
Nieves201390,Golan:2013jtj,Megias2014}. 
Multinucleon mechanisms, 2p-2h excitations and meson-exchange currents (MEC), and also RPA corrections appear essential to describe the data.
Models based on the IA, which make use either of a realistic spectral function obtained within a nonrelativistic framework \cite{Benhar:2010nx,Benhar:2011wy} or of a relativistic IA (RIA) \cite{Butkevich:2010cr,Butkevich:2011fu, jusz10,Maieron:2003df}, 
generally underestimate the MiniBooNE CCQE cross sections. Only the relativistic Green's function (RGF) model is able to give a good description of  the 
data \cite{Meucci:2011vd}. Although under many aspects based on the RIA, the RGF model can recover contributions of final-state channels that are not 
included in other models based on the RIA. These contributions are recovered by the imaginary part of the relativistic optical potential that is used 
in the RGF model to describe final-state interactions (FSI).

The RGF model was originally developed within a nonrelativistic \cite{Capuzzi:1991qd,Capuzzi:2004au} and then a relativistic  \cite{Meucci:2003uy,Meucci:2005pk} framework to describe FSI in the inclusive quasielastic (QE) electron scattering. 
The model was successfully tested against  electron scattering data \cite{Capuzzi:1991qd,Capuzzi:2004au,book,Meucci:2009nm,esotici2} 
and it was later extended to neutrino-nucleus scattering, both in the charged-current (CC) \cite{Meucci:2003cv,Meucci:2011pi,Meucci:2011vd,
Meucci:ant,Meucci:2013gja,PhysRevD.89.117301} and in the neutral-current (NC) 
\cite{Meucci:2011nc,PhysRevC.88.025502,PhysRevD.89.057302} sector. 
Although different, the two situations present many similar aspects and the extension to neutrino scattering of the electron scattering formalism
is straightforward.
In  the QE kinematic region, where the nuclear response to an electroweak probe is dominated by  single-nucleon scattering with direct one-nucleon emission, a reliable description of the FSI effects between the ejected nucleon and the residual nucleus is very important for the comparison with data. 
In the RGF model  FSI are described  in the inclusive scattering consistently with the exclusive scattering by the same complex optical potential (OP) and 
the components of the nuclear response are obtained in terms of matrix elements of the same type as the distorted wave impulse approximation ones of the
exclusive  $\eep$ process, but involve eigenfunctions of the OP and of its Hermitian conjugate, where the opposite sign of the imaginary part  
gives in one case an absorption and in the other case a gain of strength.   In the exclusive scattering, where only one channel is considered, the imaginary 
part gives an absorption that accounts for the flux lost to other channels. In the inclusive scattering, where all elastic and inelastic 
channels are included, the imaginary part redistributes the flux in all the channels and in the sum over all the channels the total 
flux is conserved. 
More details on the model can be found in our previous papers \cite{Capuzzi:1991qd,Meucci:2003uy,Meucci:2003cv,Capuzzi:2004au,Meucci:2005pk}.

In other approaches based on the RIA, FSI are included in the emitted 
nucleon state with real potentials, either retaining only the real part of the 
relativistic energy-dependent complex OP, or using 
distorted waves obtained with the same relativistic 
energy-independent mean-field potential  considered in describing the initial nucleon 
state (RMF) \cite{Maieron:2003df,Caballero:2005sn,minerva-juan}. 
In the relativistic plane-wave impulse approximation (RPWIA) FSI are neglected.

The results of these different descriptions of FSI have been compared in~\cite{Meucci:2009nm} 
for the inclusive QE electron scattering, in \cite{Meucci:2011pi} for 
the CCQE neutrino scattering, and in \cite{Meucci:2011vd,PhysRevC.88.025502} 
with the CCQE and NC elastic  MiniBooNE data.
Both RGF and RMF models can describe successfully electron scattering data and their related scaling
functions. Both models are able to provide a satisfactory description of 
the CCQE MINER$\nu$A data \cite{minerva-juan,PhysRevD.89.117301}. In the case of the MiniBooNE CCQE data, both models reproduce the 
shape of the experimental cross sections, but only the RGF gives cross sections 
of the same magnitude as the experimental ones without the need to 
increase the world average value of the axial mass $M_A$~\cite{Meucci:2011vd,Meucci:ant}. 
The larger RGF cross sections are due to the translation to the inclusive strength of the overall effect of inelastic 
channels, including rescattering and some multinucleon contributions, that  are recovered in the model by the imaginary part of the
relativistic OP and that are not included in the RMF and in other models based on the IA.

The optical potential is a powerful tool to recover and include important contributions. The availability of phenomenological relativistic OP's, obtained through a fit of elastic proton-nucleus scattering data,  is essential to make RGF calculations feasible, but  the use of a phenomenological OP does not allow us to disentangle the role of 
a specific inelastic channel and can therefore introduce uncertainties and ambiguities in the interpretation of the RGF results. 
The imaginary part can recover, to some extent, contributions beyond direct one-nucleon emission, such as, for instance,  rescattering of the outgoing nucleon and some multinucleon processes, which can be included in CCQE measurements, but the RGF model is based on the use of a one-body nuclear current and does not contain MEC mechanisms that in other models have been found to be significant.  On the other hand, the imaginary part of the OP can include pion-absorption and pion-emission processes, that should have already been subtracted in the MiniBooNE analysis.
It has been written in \cite{nieves:2014} that the good agreement of the RGF results with the MiniBooNE data ``should be interpreted with care'' and that  ``it would be 
very interesting to confront the RGF results with the fully CC-inclusive data'' \cite{nieves:2014}. The comparison with the fully CC-inclusive data, which include also pion production, is the motivation of the present paper.

The T2K collaboration has recently published \cite{PhysRevD.87.092003} new results on the 
CC-inclusive double differential cross section on $^{12}$C, which includes also pion production, and of the CCQE cross section \cite{Abe:2014iza}. The T2K $\nu_{\mu}$ energy range is the same as for MiniBooNE, the  beam peaks at $\sim 600$ MeV, similar to 
that of MiniBooNE, but it is significantly narrower and receives almost neglible contributions for energies larger 
than  $1$ GeV.
In view of these differences, the analysis of T2K data is another useful and 
independent test for a theoretical description.

In this paper we compare the results of  the  RGF model with the CC-inclusive $\nu_{\mu}$ and $\nu_e$  
 T2K cross sections on $^{12}$C \cite{PhysRevD.87.092003,t2k-nuedata,Abe:2014agb,Abe:2014iza}. 
As a first step, we compare our results with the CCQE $\nu_{\mu}-^{12}$C cross sections measured by  the MiniBooNE and T2K 
collaborations, for which there is  consistency between the two experiments within the current statistical and systematic 
uncertainties \cite{Abe:2014iza}.
Then, we consider the  flux-averaged CC-inclusive $\nu_{\mu}$ and $\nu_e$  differential cross sections from T2K with the aim to investigate whether the effects of the  inelastic channels recovered in the RGF model by the relativistic OP give or do not give enough strength  to reproduce these data.

I

 \section{Results } 
 \label{results}

In all the calculations presented in this work 
we have adopted the standard value for the nucleon axial mass 
$M_A = 1.03$ GeV$/c^2$. The bound nucleon states 
are taken as self-consistent Dirac-Hartree solutions derived 
within a relativistic mean field approach using a Lagrangian containing 
$\sigma$, $\omega$, and $\rho$ mesons 
\cite{Serot:1984ey,Rein:1989,Rign:1996,Lalazissis:1996rd,Serot:1997xg}. 
We have used two different parametrizations for the relativistic OP of $^{12}$C that is adopted in 
our RGF calculations: the
Energy-Dependent and A-Independent EDAI
(where the $E$ represents the energy and the $A$ the atomic number) 
 OP of \cite{Cooper:1993nx}, 
and  the more recent Democratic (DEM)
phenomenological OP of \cite{Cooper:2009}. The
EDAI OP is a single-nucleus parametrization, which is constructed
to better reproduce the elastic proton-$^{12}$C phenomenology, whereas 
 the DEM parametrization is a global parametrization, which
 depends on the atomic number $A$ and is obtained through 
a fit to more than 200 data sets of elastic proton-nucleus scattering data 
on a wide range of nuclei  and that is not limited to doubly closed shell nuclei.
In comparison with electron scattering data, the DEM parametrization produces 
in general good results for doubly magic nuclei and less good but still 
acceptable results for nuclei with a number of nucleons far from the magic 
numbers \cite{esotici2,PhysRevC.89.034604}.

\begin{figure}[tb]
\begin{center}
\includegraphics[scale=0.44]{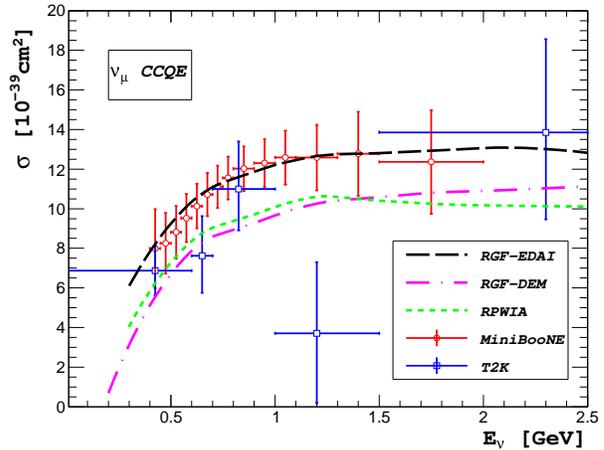} 
\end{center}
\caption{ Total  CCQE $\nu_{\mu}$-$^{12}$C cross sections per target neutron versus the neutrino energy. The experimental data are from MiniBooNE \cite{miniboone} and T2K 
\cite{Abe:2014iza}.
\label{figurat2kcstot} }
	\end{figure}
\begin{figure*}[tb]
\begin{center}
%
\includegraphics[scale=0.44]{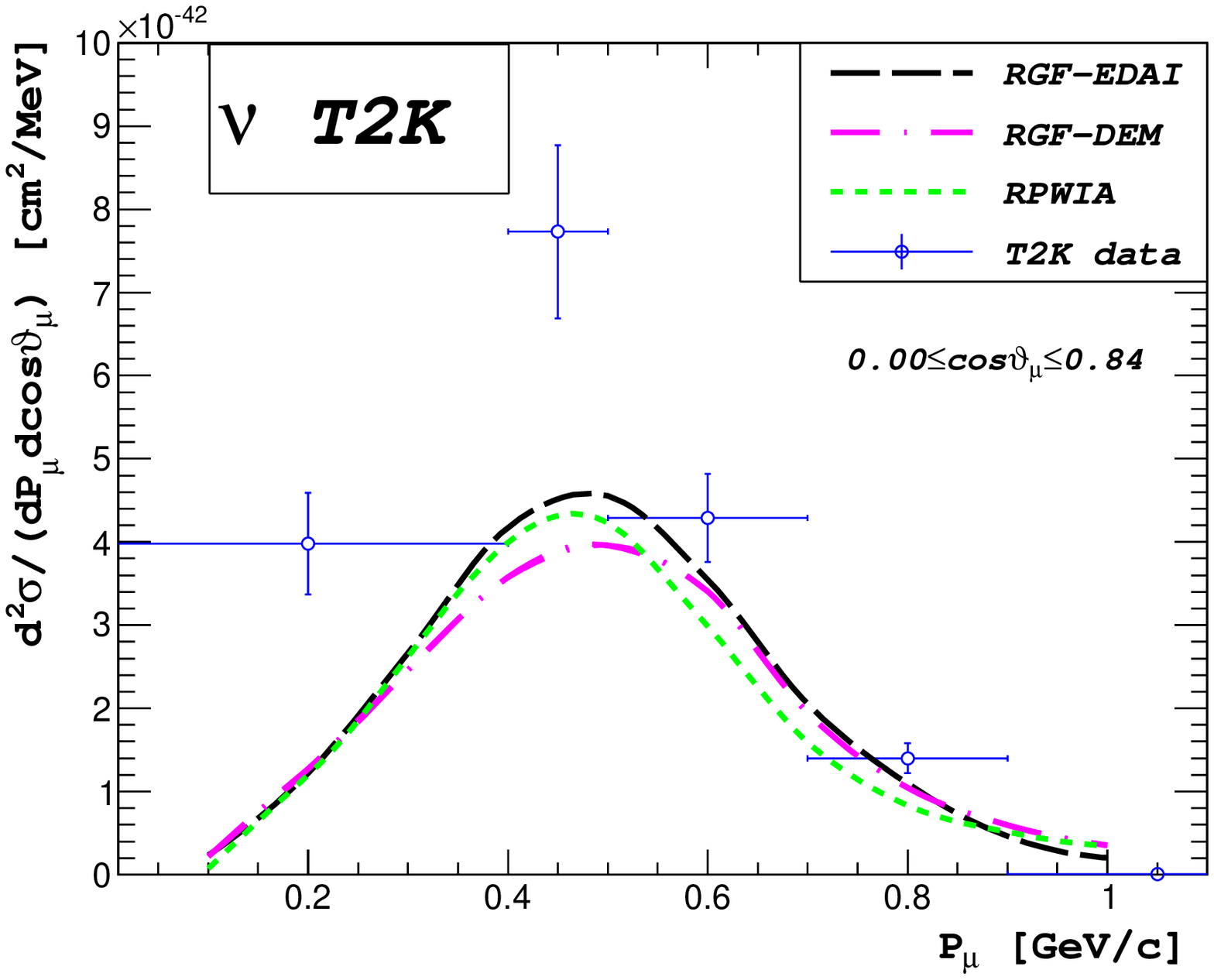} 
\includegraphics[scale=0.44]{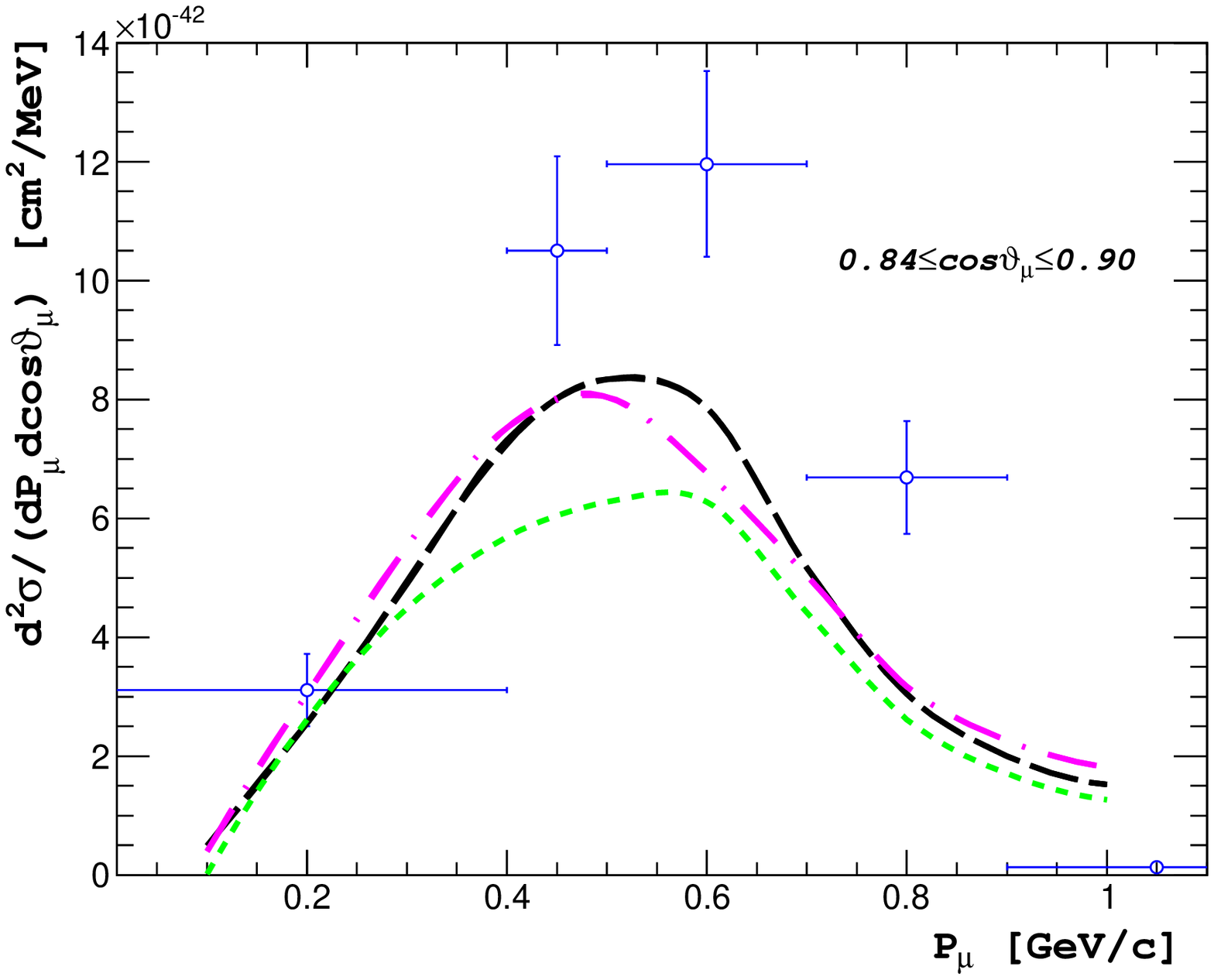} \\
\includegraphics[scale=0.44]{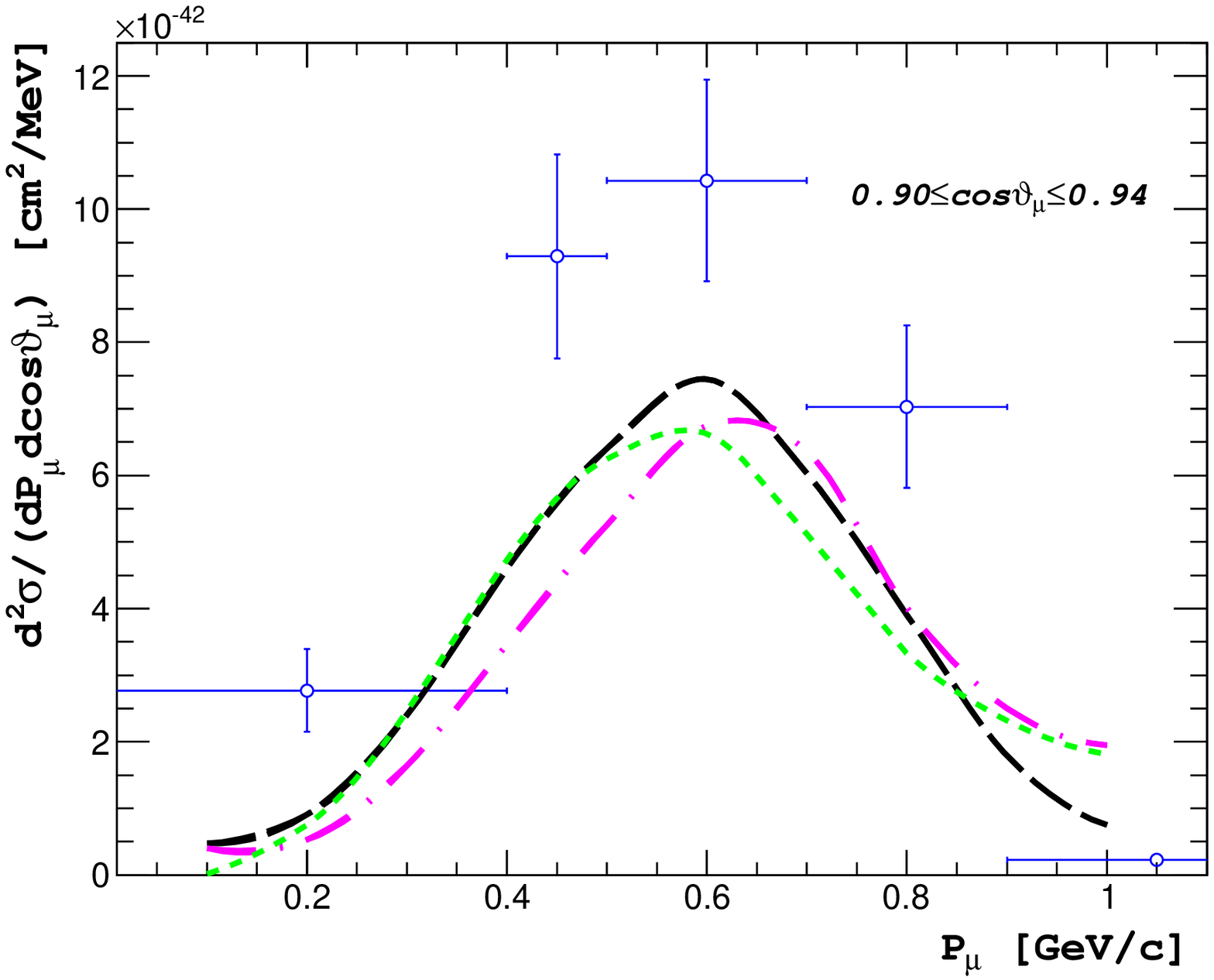} 
\includegraphics[scale=0.44]{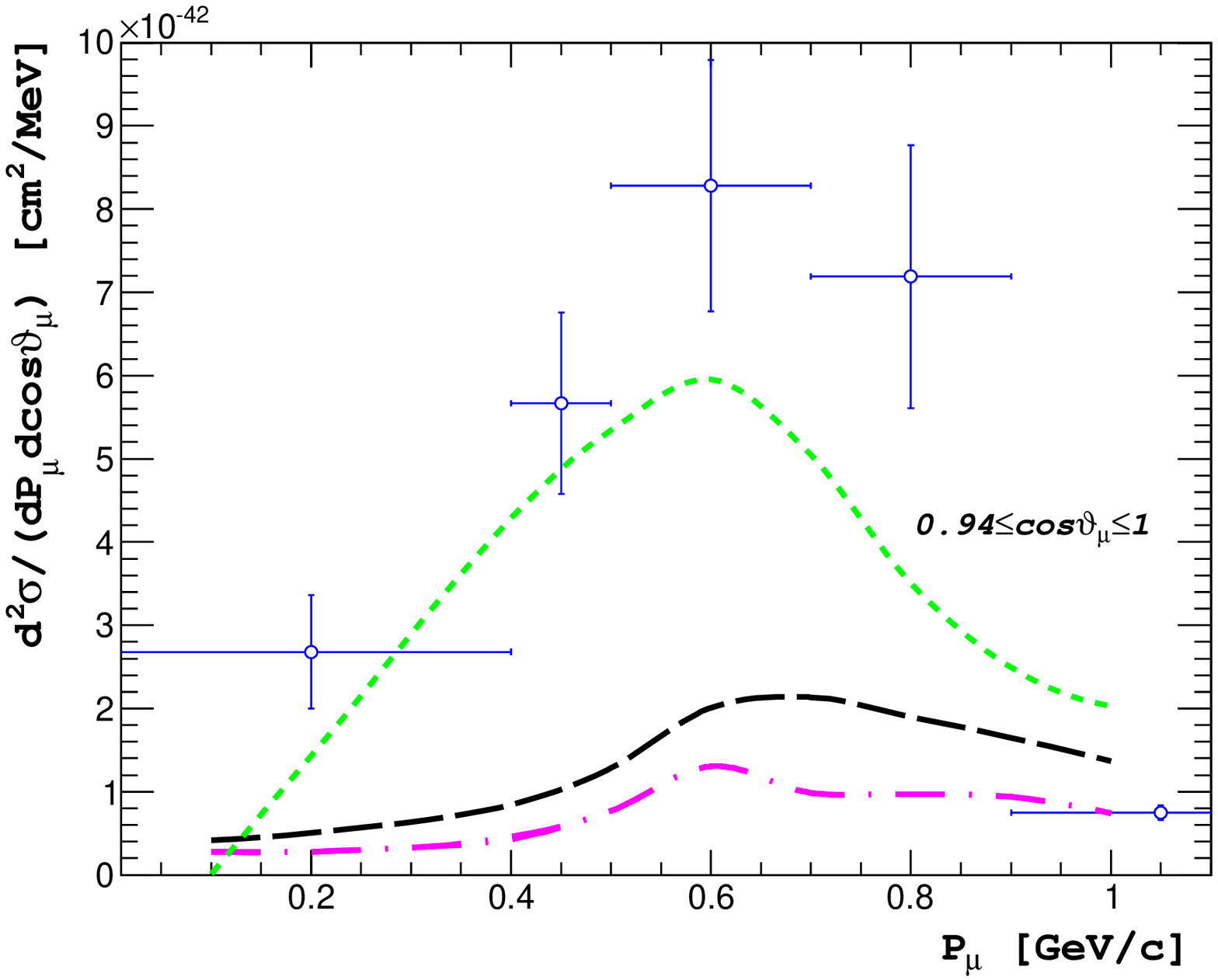} 
\end{center}
\caption{ Flux-averaged CC-inclusive  double differential $\nu_{\mu}$-$^{12}$C cross sections 
per target nucleon as a function of the muon momentum. The data are from T2K \cite{PhysRevD.87.092003}. 
\label{figurat2k} 
}
	\end{figure*}

\begin{figure}[h]
\begin{center}
\includegraphics[scale=0.44]{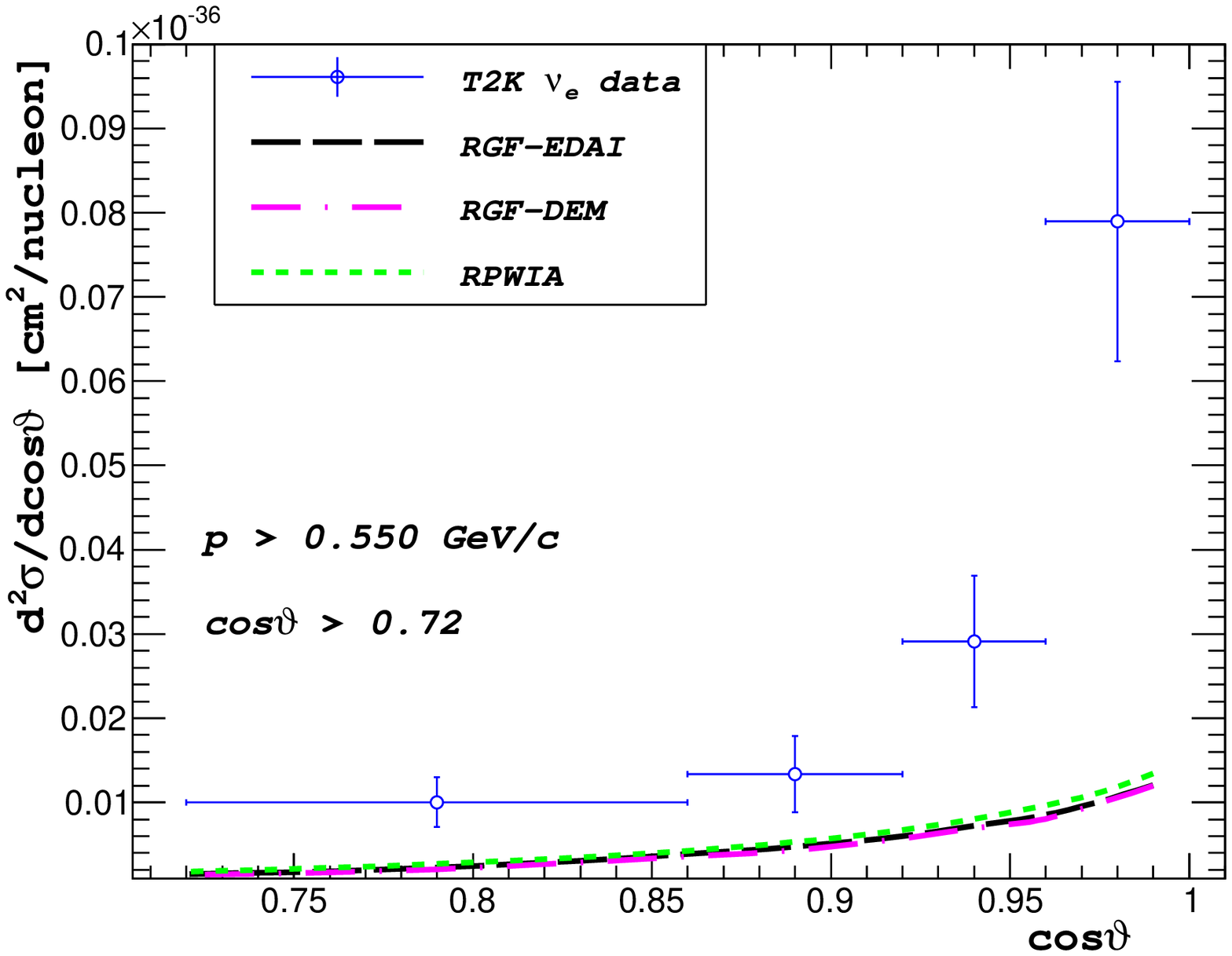} 
\end{center}
\caption{ Flux-averaged CC-inclusive $\nu_{e}$-$^{12}$C differential cross section per nucleon as a function of  $\cos{\theta}$. Only electrons corresponding to
$p > 0.550$ GeV/c and $\cos{\theta} > 0.72$ are considered \cite{Abe:2014agb}. 
The T2K data can be found
at \cite{t2k-nuedata}.
\label{figurat2knuecos} 
}
	\end{figure}
\begin{figure}[h]
\begin{center}
\includegraphics[scale=0.44]{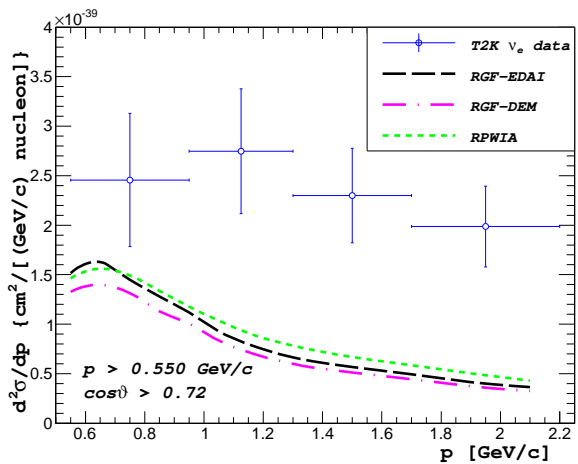} 
\end{center}
\caption{ Flux-averaged CC-inclusive  $\nu_e$-$^{12}$C  differential cross section 
per target nucleon as a function of the electron momentum. Only electrons corresponding to
$p > 0.550$ GeV/c and $\cos{\theta} > 0.72$ are considered \cite{Abe:2014agb}. 
The T2K data can be found at \cite{t2k-nuedata}.
\label{figurat2knuepe} 
}
	\end{figure}


In Fig. \ref{figurat2kcstot} we show our  calculated CCQE $\nu_{\mu}$-$^{12}$C cross sections per target neutron as a function
of the neutrino energy compared  with the MiniBooNE \cite{miniboone} and T2K data \cite{Abe:2014iza}. Although the results of these two independent measurements  
 can be consistently
compared in the entire range of energies, with the only exception of the T2K datum 
in the energy bin $1-1.5$ GeV,  we observe that the average magnitude of the MiniBooNE
 dataset is larger than that of the T2K one.
The differences between the RGF-EDAI and RGF-DEM results are sizable. These differences are due to the different imaginary parts of the 
two OPs, particularly for the energies considered in kinematics with the lowest scattering angles and the largest kinetic energies of the 
muon \cite{Meucci:2011vd}.
The RGF-EDAI cross section is larger than the RGF-DEM one, in better agreement with the MiniBooNE data and in agreement with both 
MiniBooNE and T2K cross sections within the error bars in the entire energy range of the data. 
The RGF-DEM cross section underpredicts the MiniBooNE data at low $E_{\nu}$ and  it is in
better agreement with the T2K data. 
The RPWIA cross section, which  is also shown in the figure for a comparison, is similar to the RGF-DEM one.
We note that other models based on the IA give in general results somewhat lower than the RPWIA one and therefore lower than 
the data.

In Fig. \ref{figurat2k}  we present the  CC-inclusive double differential $\nu_{\mu}$-$^{12}$C cross section 
 $d^2\sigma/(dP_{\mu} d\cos\vartheta_{\mu})$ as a function of  the outgoing muon momentum transfer $P_\mu$ for four different bins
 in  the scattering angle. The calculated cross sections are 
flux-averaged over the T2K $\nu_{\mu}$ flux \cite{PhysRevD.87.012001} and compared 
with the experimental data of \cite{PhysRevD.87.092003}. 

The RPWIA results  in Fig. \ref{figurat2k} are approximately $50\%$ lower than the data. Also the RGF results underestimate the data. 
Both  RGF-EDAI and RGF-DEM cross sections are generally lower than the data, although within the error bars for low values of $P_{\mu}$ and
large angular bins. 
A satisfactory agreement with the data is obtained with the model of \cite{Martini:2014dqa}, which includes np-nh excitations and single-pion production. 
In the RGF model the imaginary part of the OP can include the excitation of multinucleon channels. We cannot exclude that 
it can contain some  contribution  due to pion emission, we cannot 
disentangle and evaluate  the relevance of this contribution, but in any case the results in Fig. \ref{figurat2k} indicate  that this is not 
enough to reproduce CC-inclusive data.        

We note that for the most forward angular bin $(0.94 \leq \cos\vartheta_{\mu} \leq 1.00)$ the RGF results 
are significantly smaller than the RPWIA ones. 
In this small bin the transverse and charge-isovector contributions are suppressed and 
the longitudinal response gives the main contribution to the cross section. 
In addition, it has been shown in \cite{Amaro:2010sd} that  models based on quasi-free scattering cannot describe
properly this kinematic situation where $\sim 1/2$ of the total cross section arises from
excitation energies below $\sim 50$ MeV. 
If we consider that in the RGF calculations collective effects are neglected, it  
is not surprising that our results are significantly lower than the data in 
the forward angular bin.

In Fig. \ref{figurat2knuecos} and Fig. \ref{figurat2knuepe} the calculated CC-inclusive $\nu_{e}$-$^{12}$C  differential 
cross sections are displayed as a function of the electron momentum and scattering angle, respectively,  and compared with the 
T2K data of \cite{Abe:2014agb}.
For sake of simplicity the calculations have been performed only for the reduced 
phase-space $($momentum $ > 550$ MeV$/c$ 
and $\cos\vartheta > 0.72)$ and not for the full phase-space of T2K. 
The  T2K $\nu_{e}$ beam peaks at $\sim 500$ MeV and, in contrast to the $\nu_{\mu}$ one,
extends to energies larger than  $1$ GeV.
Thus the CC $\nu_{e}$ cross section at T2K may receive also higher-order
contributions like two-pion production.
Also in this case our calculations are significantly lower than the data. 
The RPWIA and the two RGF results  are very similar in Fig. \ref{figurat2knuecos}, where the three curves practically overlap, 
while in Fig. \ref{figurat2knuepe} the differences are small but visible.
The fact that collective effects are not included in our model is one of the reasons of the large underestimation of the experimental data
at smaller angle in Fig.  \ref{figurat2knuecos}.

\section{Conclusions}

In this paper we have compared the predictions of the RGF model with the
CCQE and CC-inclusive $\nu_{\mu}$ and $\nu_e$ scattering  T2K data. 
The RGF model is able to give a satisfactory description of inclusive QE electron 
scattering cross sections and of the  CCQE MiniBooNE and MINER$\nu$A data, both for $\nu_{\mu}$ and $\bar{\nu}_{\mu}$ scattering, 
 without the need to increase the standard value of the axial mass.

The RGF results are usually larger than the results of other models based on 
the IA.  In the RGF model FSI are described using a complex energy-dependent relativistic 
OP whose imaginary part includes the overall effect of the inelastic 
channels, which give different contributions at different energies, and makes 
the RGF results sensitive to the kinematic conditions of the calculations. 
With a complex OP the model can include all the available 
final-state channels and not only direct one-nucleon emission processes.
The important role of contributions other than direct 
one-nucleon emission has been confirmed by different and independent models in the case of
CCQE MiniBooNE cross sections, but  the same  
conclusion is doubtful in the case of MINER$\nu$A data.

The RGF model can include contributions of final-state channels like, e.g., rescattering processes of the nucleon in its way out of the nucleus,  
non-nucleonic $\Delta$ excitations, which may arise during nucleon propagation, with or without real-pion production, and also some multinucleon processes.
These contributions are not incorporated explicitely in the model with a microscopic calculation: they can be recovered, to some extent, 
at a phenomenological level by the imaginary part of the phenomenological OP which is adopted in the RGF calculation. 
The use of a  phenomenological OP does not allow us to disentangle and evaluate the role of a specific reaction process. 
Different available parametrizations of the phenomenological relativistic OP can introduce uncertainties in the predictions 
of the model. 
 The determination of a theoretical OP, which fulfills the dispersion relations in the whole energy region 
of interest, would be very useful to reduce the theoretical uncertainties.


In the RGF model the nuclear response is written in terms of the single-particle optical-model Green's function. This result is obtained retaining 
only the one-body part of the nuclear current. The inclusion of  two-body MEC would require an extended model 
based on the two-particle Green's function, whose evaluation represents a very hard task. 
   
The imaginary part of the OP can include pion-production processes that should have already been subtracted in the analysis 
of CCQE data. 
In this paper we have shown that the fully CC-inclusive T2K cross sections, which include pion production, are clearly underestimated by the RGF calculations. 
Even if we cannot disentangle the pion-production contribution that can be included in the phenomenological OP, this is not enough to reproduce the CC-inclusive T2K data. 
If we consider that the RGF model was developed to describe FSI in the inclusive QE scattering and that it is able to give a reasonably good agreement with QE electron and CCQE neutrino-scattering data,  
the present comparison with T2K data can be interpreted as an indication that the pion-production channel gives only a minor contribution to the RGF results.  

The full and explicit inclusion of multinucleon channels is required 
to successfully reproduce CC-inclusive data. Other models, which explicitly include multinucleon emission channels, 
 obtain a satisfactory agreement  with the T2K data \cite{Martini:2014dqa}. 

\begin{acknowledgements}

We thank Marco Martini for interesting and useful discussions.

\end{acknowledgements}


\begin{thebibliography}{56}%
\makeatletter
\providecommand \@ifxundefined [1]{%
 \@ifx{#1\undefined}
}%
\providecommand \@ifnum [1]{%
 \ifnum #1\expandafter \@firstoftwo
 \else \expandafter \@secondoftwo
 \fi
}%
\providecommand \@ifx [1]{%
 \ifx #1\expandafter \@firstoftwo
 \else \expandafter \@secondoftwo
 \fi
}%
\providecommand \natexlab [1]{#1}%
\providecommand \enquote  [1]{``#1''}%
\providecommand \bibnamefont  [1]{#1}%
\providecommand \bibfnamefont [1]{#1}%
\providecommand \citenamefont [1]{#1}%
\providecommand \href@noop [0]{\@secondoftwo}%
\providecommand \href [0]{\begingroup \@sanitize@url \@href}%
\providecommand \@href[1]{\@@startlink{#1}\@@href}%
\providecommand \@@href[1]{\endgroup#1\@@endlink}%
\providecommand \@sanitize@url [0]{\catcode `\\12\catcode `\$12\catcode
  `\&12\catcode `\#12\catcode `\^12\catcode `\_12\catcode `\%12\relax}%
\providecommand \@@startlink[1]{}%
\providecommand \@@endlink[0]{}%
\providecommand \url  [0]{\begingroup\@sanitize@url \@url }%
\providecommand \@url [1]{\endgroup\@href {#1}{\urlprefix }}%
\providecommand \urlprefix  [0]{URL }%
\providecommand \Eprint [0]{\href }%
\providecommand \doibase [0]{http://dx.doi.org/}%
\providecommand \selectlanguage [0]{\@gobble}%
\providecommand \bibinfo  [0]{\@secondoftwo}%
\providecommand \bibfield  [0]{\@secondoftwo}%
\providecommand \translation [1]{[#1]}%
\providecommand \BibitemOpen [0]{}%
\providecommand \bibitemStop [0]{}%
\providecommand \bibitemNoStop [0]{.\EOS\space}%
\providecommand \EOS [0]{\spacefactor3000\relax}%
\providecommand \BibitemShut  [1]{\csname bibitem#1\endcsname}%
\let\auto@bib@innerbib\@empty
\bibitem [{\citenamefont {Formaggio}\ and\ \citenamefont
  {Zeller}(2012)}]{RevModPhys.84.1307}%
  \BibitemOpen
  \bibfield  {author} {\bibinfo {author} {\bibfnamefont {J.~A.}\ \bibnamefont
  {Formaggio}}\ and\ \bibinfo {author} {\bibfnamefont {G.~P.}\ \bibnamefont
  {Zeller}},\ }\href {\doibase 10.1103/RevModPhys.84.1307} {\bibfield
  {journal} {\bibinfo  {journal} {Rev. Mod. Phys.}\ }\textbf {\bibinfo {volume}
  {84}},\ \bibinfo {pages} {1307} (\bibinfo {year} {2012})}\BibitemShut
  {NoStop}%
\bibitem [{\citenamefont {Morfin}\ \emph {et~al.}(2012)\citenamefont {Morfin},
  \citenamefont {Nieves},\ and\ \citenamefont {Sobczyk}}]{Morfin:2012kn}%
  \BibitemOpen
  \bibfield  {author} {\bibinfo {author} {\bibfnamefont {J.~G.}\ \bibnamefont
  {Morfin}}, \bibinfo {author} {\bibfnamefont {J.}~\bibnamefont {Nieves}}, \
  and\ \bibinfo {author} {\bibfnamefont {J.~T.}\ \bibnamefont {Sobczyk}},\
  }\href {\doibase 10.1155/2012/934597} {\bibfield  {journal} {\bibinfo
  {journal} {Adv. High Energy Phys.}\ }\textbf {\bibinfo {volume} {2012}},\
  \bibinfo {pages} {934597} (\bibinfo {year} {2012})}\BibitemShut {NoStop}%
\bibitem [{\citenamefont {Alvarez-Ruso}\ \emph {et~al.}(2014)\citenamefont
  {Alvarez-Ruso}, \citenamefont {Hayato},\ and\ \citenamefont
  {Nieves}}]{nieves:2014}%
  \BibitemOpen
  \bibfield  {author} {\bibinfo {author} {\bibfnamefont {L.}~\bibnamefont
  {Alvarez-Ruso}}, \bibinfo {author} {\bibfnamefont {Y.}~\bibnamefont
  {Hayato}}, \ and\ \bibinfo {author} {\bibfnamefont {J.}~\bibnamefont
  {Nieves}},\ }\href {\doibase 10.1088/1367-2630/16/7/075015} {\bibfield
  {journal} {\bibinfo  {journal} {{New Journal of Physics}}\ }\textbf {\bibinfo
  {volume} {16}},\ \bibinfo {pages} {075015} (\bibinfo {year}
  {2014})}\BibitemShut {NoStop}%
\bibitem [{\citenamefont {Aguilar-Arevalo}\ \emph {et~al.}(2010)\citenamefont
  {Aguilar-Arevalo} \emph {et~al.}}]{miniboone}%
  \BibitemOpen
  \bibfield  {author} {\bibinfo {author} {\bibfnamefont {A.~A.}\ \bibnamefont
  {Aguilar-Arevalo}} \emph {et~al.} (\bibinfo {collaboration} {MiniBooNE
  Collaboration}),\ }\href@noop {} {\bibfield  {journal} {\bibinfo  {journal}
  {Phys. Rev. D}\ }\textbf {\bibinfo {volume} {81}},\ \bibinfo {pages} {092005}
  (\bibinfo {year} {2010})}\BibitemShut {NoStop}%
\bibitem [{\citenamefont {Aguilar-Arevalo}\ \emph {et~al.}(2013)\citenamefont
  {Aguilar-Arevalo} \emph {et~al.}}]{miniboone-ant}%
  \BibitemOpen
  \bibfield  {author} {\bibinfo {author} {\bibfnamefont {A.~A.}\ \bibnamefont
  {Aguilar-Arevalo}} \emph {et~al.} (\bibinfo {collaboration} {MiniBooNE
  Collaboration}),\ }\href {\doibase 10.1103/PhysRevD.88.032001} {\bibfield
  {journal} {\bibinfo  {journal} {Phys. Rev. D}\ }\textbf {\bibinfo {volume}
  {88}},\ \bibinfo {pages} {032001} (\bibinfo {year} {2013})}\BibitemShut
  {NoStop}%
\bibitem [{\citenamefont {Leitner}\ \emph {et~al.}(2009)\citenamefont
  {Leitner}, \citenamefont {Buss}, \citenamefont {Alvarez-Ruso},\ and\
  \citenamefont {Mosel}}]{PhysRevC.79.034601}%
  \BibitemOpen
  \bibfield  {author} {\bibinfo {author} {\bibfnamefont {T.}~\bibnamefont
  {Leitner}}, \bibinfo {author} {\bibfnamefont {O.}~\bibnamefont {Buss}},
  \bibinfo {author} {\bibfnamefont {L.}~\bibnamefont {Alvarez-Ruso}}, \ and\
  \bibinfo {author} {\bibfnamefont {U.}~\bibnamefont {Mosel}},\ }\href
  {\doibase 10.1103/PhysRevC.79.034601} {\bibfield  {journal} {\bibinfo
  {journal} {Phys. Rev. C}\ }\textbf {\bibinfo {volume} {79}},\ \bibinfo
  {pages} {034601} (\bibinfo {year} {2009})}\BibitemShut {NoStop}%
\bibitem [{\citenamefont {Martini}\ \emph {et~al.}(2009)\citenamefont
  {Martini}, \citenamefont {Ericson}, \citenamefont {Chanfray},\ and\
  \citenamefont {Marteau}}]{Martini:2009uj}%
  \BibitemOpen
  \bibfield  {author} {\bibinfo {author} {\bibfnamefont {M.}~\bibnamefont
  {Martini}}, \bibinfo {author} {\bibfnamefont {M.}~\bibnamefont {Ericson}},
  \bibinfo {author} {\bibfnamefont {G.}~\bibnamefont {Chanfray}}, \ and\
  \bibinfo {author} {\bibfnamefont {J.}~\bibnamefont {Marteau}},\ }\href
  {\doibase 10.1103/PhysRevC.80.065501} {\bibfield  {journal} {\bibinfo
  {journal} {Phys. Rev. C}\ }\textbf {\bibinfo {volume} {80}},\ \bibinfo
  {pages} {065501} (\bibinfo {year} {2009})}\BibitemShut {NoStop}%
\bibitem [{\citenamefont {Martini}\ \emph {et~al.}(2010)\citenamefont
  {Martini}, \citenamefont {Ericson}, \citenamefont {Chanfray},\ and\
  \citenamefont {Marteau}}]{Martini:2010ex}%
  \BibitemOpen
  \bibfield  {author} {\bibinfo {author} {\bibfnamefont {M.}~\bibnamefont
  {Martini}}, \bibinfo {author} {\bibfnamefont {M.}~\bibnamefont {Ericson}},
  \bibinfo {author} {\bibfnamefont {G.}~\bibnamefont {Chanfray}}, \ and\
  \bibinfo {author} {\bibfnamefont {J.}~\bibnamefont {Marteau}},\ }\href
  {\doibase 10.1103/PhysRevC.81.045502} {\bibfield  {journal} {\bibinfo
  {journal} {Phys. Rev. C}\ }\textbf {\bibinfo {volume} {81}},\ \bibinfo
  {pages} {045502} (\bibinfo {year} {2010})}\BibitemShut {NoStop}%
\bibitem [{\citenamefont {Martini}\ and\ \citenamefont
  {Ericson}(2013)}]{Martini:2013sha}%
  \BibitemOpen
  \bibfield  {author} {\bibinfo {author} {\bibfnamefont {M.}~\bibnamefont
  {Martini}}\ and\ \bibinfo {author} {\bibfnamefont {M.}~\bibnamefont
  {Ericson}},\ }\href {\doibase 10.1103/PhysRevC.87.065501} {\bibfield
  {journal} {\bibinfo  {journal} {Phys. Rev. C}\ }\textbf {\bibinfo {volume}
  {87}},\ \bibinfo {pages} {065501} (\bibinfo {year} {2013})}\BibitemShut
  {NoStop}%
\bibitem [{\citenamefont {Leitner}\ and\ \citenamefont
  {Mosel}(2010)}]{Leitner:2010kp}%
  \BibitemOpen
  \bibfield  {author} {\bibinfo {author} {\bibfnamefont {T.}~\bibnamefont
  {Leitner}}\ and\ \bibinfo {author} {\bibfnamefont {U.}~\bibnamefont
  {Mosel}},\ }\href {\doibase 10.1103/PhysRevC.81.064614} {\bibfield  {journal}
  {\bibinfo  {journal} {Phys. Rev. C}\ }\textbf {\bibinfo {volume} {81}},\
  \bibinfo {pages} {064614} (\bibinfo {year} {2010})}\BibitemShut {NoStop}%
\bibitem [{\citenamefont {Ankowski}\ and\ \citenamefont
  {Benhar}(2011)}]{PhysRevC.83.054616}%
  \BibitemOpen
  \bibfield  {author} {\bibinfo {author} {\bibfnamefont {A.~M.}\ \bibnamefont
  {Ankowski}}\ and\ \bibinfo {author} {\bibfnamefont {O.}~\bibnamefont
  {Benhar}},\ }\href {\doibase 10.1103/PhysRevC.83.054616} {\bibfield
  {journal} {\bibinfo  {journal} {Phys. Rev. C}\ }\textbf {\bibinfo {volume}
  {83}},\ \bibinfo {pages} {054616} (\bibinfo {year} {2011})}\BibitemShut
  {NoStop}%
\bibitem [{\citenamefont {Fernandez~Martinez}\ and\ \citenamefont
  {Meloni}(2011)}]{FernandezMartinez2011477}%
  \BibitemOpen
  \bibfield  {author} {\bibinfo {author} {\bibfnamefont {E.}~\bibnamefont
  {Fernandez~Martinez}}\ and\ \bibinfo {author} {\bibfnamefont
  {D.}~\bibnamefont {Meloni}},\ }\href {\doibase
  10.1016/j.physletb.2011.02.043} {\bibfield  {journal} {\bibinfo  {journal}
  {Physics Letters B}\ }\textbf {\bibinfo {volume} {697}},\ \bibinfo {pages}
  {477 } (\bibinfo {year} {2011})}\BibitemShut {NoStop}%
\bibitem [{\citenamefont {Amaro}\ \emph {et~al.}(2012)\citenamefont {Amaro},
  \citenamefont {Barbaro}, \citenamefont {Caballero},\ and\ \citenamefont
  {Donnelly}}]{AmaroAntSusa}%
  \BibitemOpen
  \bibfield  {author} {\bibinfo {author} {\bibfnamefont {J.~E.}\ \bibnamefont
  {Amaro}}, \bibinfo {author} {\bibfnamefont {M.~B.}\ \bibnamefont {Barbaro}},
  \bibinfo {author} {\bibfnamefont {J.~A.}\ \bibnamefont {Caballero}}, \ and\
  \bibinfo {author} {\bibfnamefont {T.~W.}\ \bibnamefont {Donnelly}},\ }\href
  {\doibase 10.1103/PhysRevLett.108.152501} {\bibfield  {journal} {\bibinfo
  {journal} {Phys. Rev. Lett.}\ }\textbf {\bibinfo {volume} {108}},\ \bibinfo
  {pages} {152501} (\bibinfo {year} {2012})}\BibitemShut {NoStop}%
\bibitem [{\citenamefont {Amaro}\ \emph
  {et~al.}(2011{\natexlab{a}})\citenamefont {Amaro}, \citenamefont {Barbaro},
  \citenamefont {Caballero}, \citenamefont {Donnelly},\ and\ \citenamefont
  {Ud\'{\i}as}}]{Amaro:2011qb}%
  \BibitemOpen
  \bibfield  {author} {\bibinfo {author} {\bibfnamefont {J.~E.}\ \bibnamefont
  {Amaro}}, \bibinfo {author} {\bibfnamefont {M.~B.}\ \bibnamefont {Barbaro}},
  \bibinfo {author} {\bibfnamefont {J.~A.}\ \bibnamefont {Caballero}}, \bibinfo
  {author} {\bibfnamefont {T.~W.}\ \bibnamefont {Donnelly}}, \ and\ \bibinfo
  {author} {\bibfnamefont {J.~M.}\ \bibnamefont {Ud\'{\i}as}},\ }\href
  {\doibase 10.1103/PhysRevD.84.033004} {\bibfield  {journal} {\bibinfo
  {journal} {Phys. Rev. D}\ }\textbf {\bibinfo {volume} {84}},\ \bibinfo
  {pages} {033004} (\bibinfo {year} {2011}{\natexlab{a}})}\BibitemShut
  {NoStop}%
\bibitem [{\citenamefont {Nieves}\ \emph {et~al.}(2013)\citenamefont {Nieves},
  \citenamefont {Ruiz~Simo},\ and\ \citenamefont
  {Vicente~Vacas}}]{Nieves201390}%
  \BibitemOpen
  \bibfield  {author} {\bibinfo {author} {\bibfnamefont {J.}~\bibnamefont
  {Nieves}}, \bibinfo {author} {\bibfnamefont {I.}~\bibnamefont {Ruiz~Simo}}, \
  and\ \bibinfo {author} {\bibfnamefont {M.~J.}\ \bibnamefont
  {Vicente~Vacas}},\ }\href {\doibase 10.1016/j.physletb.2013.03.002}
  {\bibfield  {journal} {\bibinfo  {journal} {Physics Letters B}\ }\textbf
  {\bibinfo {volume} {721}},\ \bibinfo {pages} {90 } (\bibinfo {year}
  {2013})}\BibitemShut {NoStop}%
\bibitem [{\citenamefont {Golan}\ \emph {et~al.}(2013)\citenamefont {Golan},
  \citenamefont {Graczyk}, \citenamefont {Juszczak},\ and\ \citenamefont
  {Sobczyk}}]{Golan:2013jtj}%
  \BibitemOpen
  \bibfield  {author} {\bibinfo {author} {\bibfnamefont {T.}~\bibnamefont
  {Golan}}, \bibinfo {author} {\bibfnamefont {K.~M.}\ \bibnamefont {Graczyk}},
  \bibinfo {author} {\bibfnamefont {C.}~\bibnamefont {Juszczak}}, \ and\
  \bibinfo {author} {\bibfnamefont {J.~T.}\ \bibnamefont {Sobczyk}},\ }\href
  {\doibase 10.1103/PhysRevC.88.024612} {\bibfield  {journal} {\bibinfo
  {journal} {Phys. Rev. C}\ }\textbf {\bibinfo {volume} {88}},\ \bibinfo
  {pages} {024612} (\bibinfo {year} {2013})}\BibitemShut {NoStop}%
\bibitem [{\citenamefont {Megias}\ \emph
  {et~al.}(2014{\natexlab{a}})\citenamefont {Megias}, \citenamefont {Donnelly},
  \citenamefont {Moreno}, \citenamefont {Williamson}, \citenamefont
  {Caballero}, \citenamefont {Gonz\'alez-Jim\'enez}, \citenamefont {De~Pace},
  \citenamefont {Barbaro}, \citenamefont {Alberico}, \citenamefont {Nardi},\
  and\ \citenamefont {Amaro}}]{Megias2014}%
  \BibitemOpen
  \bibfield  {author} {\bibinfo {author} {\bibfnamefont {G.~D.}\ \bibnamefont
  {Megias}}, \bibinfo {author} {\bibfnamefont {T.~W.}\ \bibnamefont
  {Donnelly}}, \bibinfo {author} {\bibfnamefont {O.}~\bibnamefont {Moreno}},
  \bibinfo {author} {\bibfnamefont {C.~F.}\ \bibnamefont {Williamson}},
  \bibinfo {author} {\bibfnamefont {J.~A.}\ \bibnamefont {Caballero}}, \bibinfo
  {author} {\bibfnamefont {R.}~\bibnamefont {Gonz\'alez-Jim\'enez}}, \bibinfo
  {author} {\bibfnamefont {A.}~\bibnamefont {De~Pace}}, \bibinfo {author}
  {\bibfnamefont {M.~B.}\ \bibnamefont {Barbaro}}, \bibinfo {author}
  {\bibfnamefont {W.~M.}\ \bibnamefont {Alberico}}, \bibinfo {author}
  {\bibfnamefont {M.}~\bibnamefont {Nardi}}, \ and\ \bibinfo {author}
  {\bibfnamefont {J.~E.}\ \bibnamefont {Amaro}},\ }\href@noop {} {\  (\bibinfo
  {year} {2014}{\natexlab{a}})},\ \Eprint {http://arxiv.org/abs/{1412.1822}}
  {{1412.1822} [nucl-th]} \BibitemShut {NoStop}%
\bibitem [{\citenamefont {Benhar}\ \emph {et~al.}(2010)\citenamefont {Benhar},
  \citenamefont {Coletti},\ and\ \citenamefont {Meloni}}]{Benhar:2010nx}%
  \BibitemOpen
  \bibfield  {author} {\bibinfo {author} {\bibfnamefont {O.}~\bibnamefont
  {Benhar}}, \bibinfo {author} {\bibfnamefont {P.}~\bibnamefont {Coletti}}, \
  and\ \bibinfo {author} {\bibfnamefont {D.}~\bibnamefont {Meloni}},\ }\href
  {\doibase 10.1103/PhysRevLett.105.132301} {\bibfield  {journal} {\bibinfo
  {journal} {Phys. Rev. Lett.}\ }\textbf {\bibinfo {volume} {105}},\ \bibinfo
  {pages} {132301} (\bibinfo {year} {2010})}\BibitemShut {NoStop}%
\bibitem [{\citenamefont {Benhar}\ and\ \citenamefont
  {Veneziano}(2011)}]{Benhar:2011wy}%
  \BibitemOpen
  \bibfield  {author} {\bibinfo {author} {\bibfnamefont {O.}~\bibnamefont
  {Benhar}}\ and\ \bibinfo {author} {\bibfnamefont {G.}~\bibnamefont
  {Veneziano}},\ }\href {\doibase 10.1016/j.physletb.2011.07.032} {\bibfield
  {journal} {\bibinfo  {journal} {Phys. Lett. B}\ }\textbf {\bibinfo {volume}
  {702}},\ \bibinfo {pages} {433} (\bibinfo {year} {2011})}\BibitemShut
  {NoStop}%
\bibitem [{\citenamefont {Butkevich}(2010)}]{Butkevich:2010cr}%
  \BibitemOpen
  \bibfield  {author} {\bibinfo {author} {\bibfnamefont {A.~V.}\ \bibnamefont
  {Butkevich}},\ }\href {\doibase 10.1103/PhysRevC.82.055501} {\bibfield
  {journal} {\bibinfo  {journal} {Phys. Rev. C}\ }\textbf {\bibinfo {volume}
  {82}},\ \bibinfo {pages} {055501} (\bibinfo {year} {2010})}\BibitemShut
  {NoStop}%
\bibitem [{\citenamefont {Butkevich}\ and\ \citenamefont
  {Perevalov}(2011)}]{Butkevich:2011fu}%
  \BibitemOpen
  \bibfield  {author} {\bibinfo {author} {\bibfnamefont {A.~V.}\ \bibnamefont
  {Butkevich}}\ and\ \bibinfo {author} {\bibfnamefont {D.}~\bibnamefont
  {Perevalov}},\ }\href {\doibase 10.1103/PhysRevC.84.015501} {\bibfield
  {journal} {\bibinfo  {journal} {Phys. Rev. C}\ }\textbf {\bibinfo {volume}
  {84}},\ \bibinfo {pages} {015501} (\bibinfo {year} {2011})}\BibitemShut
  {NoStop}%
\bibitem [{\citenamefont {Juszczak}\ \emph {et~al.}(2010)\citenamefont
  {Juszczak}, \citenamefont {Sobczyk},\ and\ \citenamefont {Zmuda}}]{jusz10}%
  \BibitemOpen
  \bibfield  {author} {\bibinfo {author} {\bibfnamefont {C.}~\bibnamefont
  {Juszczak}}, \bibinfo {author} {\bibfnamefont {J.~T.}\ \bibnamefont
  {Sobczyk}}, \ and\ \bibinfo {author} {\bibfnamefont {J.}~\bibnamefont
  {Zmuda}},\ }\href@noop {} {\bibfield  {journal} {\bibinfo  {journal} {Phys.
  Rev. C}\ }\textbf {\bibinfo {volume} {82}},\ \bibinfo {pages} {045502}
  (\bibinfo {year} {2010})}\BibitemShut {NoStop}%
\bibitem [{\citenamefont {Maieron}\ \emph {et~al.}(2003)\citenamefont
  {Maieron}, \citenamefont {Martinez}, \citenamefont {Caballero},\ and\
  \citenamefont {Ud\'{\i}as}}]{Maieron:2003df}%
  \BibitemOpen
  \bibfield  {author} {\bibinfo {author} {\bibfnamefont {C.}~\bibnamefont
  {Maieron}}, \bibinfo {author} {\bibfnamefont {M.~C.}\ \bibnamefont
  {Martinez}}, \bibinfo {author} {\bibfnamefont {J.~A.}\ \bibnamefont
  {Caballero}}, \ and\ \bibinfo {author} {\bibfnamefont {J.~M.}\ \bibnamefont
  {Ud\'{\i}as}},\ }\href {\doibase 10.1103/PhysRevC.68.048501} {\bibfield
  {journal} {\bibinfo  {journal} {Phys. Rev. C}\ }\textbf {\bibinfo {volume}
  {68}},\ \bibinfo {pages} {048501} (\bibinfo {year} {2003})}\BibitemShut
  {NoStop}%
\bibitem [{\citenamefont {Meucci}\ \emph
  {et~al.}(2011{\natexlab{a}})\citenamefont {Meucci}, \citenamefont {Barbaro},
  \citenamefont {Caballero}, \citenamefont {Giusti},\ and\ \citenamefont
  {Ud\'{\i}as}}]{Meucci:2011vd}%
  \BibitemOpen
  \bibfield  {author} {\bibinfo {author} {\bibfnamefont {A.}~\bibnamefont
  {Meucci}}, \bibinfo {author} {\bibfnamefont {M.~B.}\ \bibnamefont {Barbaro}},
  \bibinfo {author} {\bibfnamefont {J.~A.}\ \bibnamefont {Caballero}}, \bibinfo
  {author} {\bibfnamefont {C.}~\bibnamefont {Giusti}}, \ and\ \bibinfo {author}
  {\bibfnamefont {J.~M.}\ \bibnamefont {Ud\'{\i}as}},\ }\href {\doibase
  10.1103/PhysRevLett.107.172501} {\bibfield  {journal} {\bibinfo  {journal}
  {Phys. Rev. Lett.}\ }\textbf {\bibinfo {volume} {107}},\ \bibinfo {pages}
  {172501} (\bibinfo {year} {2011}{\natexlab{a}})}\BibitemShut {NoStop}%
\bibitem [{\citenamefont {Capuzzi}\ \emph {et~al.}(1991)\citenamefont
  {Capuzzi}, \citenamefont {Giusti},\ and\ \citenamefont
  {Pacati}}]{Capuzzi:1991qd}%
  \BibitemOpen
  \bibfield  {author} {\bibinfo {author} {\bibfnamefont {F.}~\bibnamefont
  {Capuzzi}}, \bibinfo {author} {\bibfnamefont {C.}~\bibnamefont {Giusti}}, \
  and\ \bibinfo {author} {\bibfnamefont {F.~D.}\ \bibnamefont {Pacati}},\
  }\href {\doibase 10.1016/0375-9474(91)90269-C} {\bibfield  {journal}
  {\bibinfo  {journal} {Nuclear Physics A}\ }\textbf {\bibinfo {volume}
  {524}},\ \bibinfo {pages} {681 } (\bibinfo {year} {1991})}\BibitemShut
  {NoStop}%
\bibitem [{\citenamefont {Capuzzi}\ \emph {et~al.}(2005)\citenamefont
  {Capuzzi}, \citenamefont {Giusti}, \citenamefont {Pacati},\ and\
  \citenamefont {Kadrev}}]{Capuzzi:2004au}%
  \BibitemOpen
  \bibfield  {author} {\bibinfo {author} {\bibfnamefont {F.}~\bibnamefont
  {Capuzzi}}, \bibinfo {author} {\bibfnamefont {C.}~\bibnamefont {Giusti}},
  \bibinfo {author} {\bibfnamefont {F.~D.}\ \bibnamefont {Pacati}}, \ and\
  \bibinfo {author} {\bibfnamefont {D.~N.}\ \bibnamefont {Kadrev}},\ }\href
  {\doibase 10.1016/j.aop.2004.12.005} {\bibfield  {journal} {\bibinfo
  {journal} {Annals of Physics (N.Y.)}\ }\textbf {\bibinfo {volume} {317}},\
  \bibinfo {pages} {492 } (\bibinfo {year} {2005})}\BibitemShut {NoStop}%
\bibitem [{\citenamefont {Meucci}\ \emph {et~al.}(2003)\citenamefont {Meucci},
  \citenamefont {Capuzzi}, \citenamefont {Giusti},\ and\ \citenamefont
  {Pacati}}]{Meucci:2003uy}%
  \BibitemOpen
  \bibfield  {author} {\bibinfo {author} {\bibfnamefont {A.}~\bibnamefont
  {Meucci}}, \bibinfo {author} {\bibfnamefont {F.}~\bibnamefont {Capuzzi}},
  \bibinfo {author} {\bibfnamefont {C.}~\bibnamefont {Giusti}}, \ and\ \bibinfo
  {author} {\bibfnamefont {F.~D.}\ \bibnamefont {Pacati}},\ }\href {\doibase
  10.1103/PhysRevC.67.054601} {\bibfield  {journal} {\bibinfo  {journal} {Phys.
  Rev. C}\ }\textbf {\bibinfo {volume} {67}},\ \bibinfo {pages} {054601}
  (\bibinfo {year} {2003})}\BibitemShut {NoStop}%
\bibitem [{\citenamefont {Meucci}\ \emph {et~al.}(2005)\citenamefont {Meucci},
  \citenamefont {Giusti},\ and\ \citenamefont {Pacati}}]{Meucci:2005pk}%
  \BibitemOpen
  \bibfield  {author} {\bibinfo {author} {\bibfnamefont {A.}~\bibnamefont
  {Meucci}}, \bibinfo {author} {\bibfnamefont {C.}~\bibnamefont {Giusti}}, \
  and\ \bibinfo {author} {\bibfnamefont {F.~D.}\ \bibnamefont {Pacati}},\
  }\href {\doibase 10.1016/j.nuclphysa.2005.04.007} {\bibfield  {journal}
  {\bibinfo  {journal} {Nuclear Physics A}\ }\textbf {\bibinfo {volume}
  {756}},\ \bibinfo {pages} {359} (\bibinfo {year} {2005})}\BibitemShut
  {NoStop}%
\bibitem [{\citenamefont {Boffi}\ \emph {et~al.}(1996)\citenamefont {Boffi},
  \citenamefont {Giusti}, \citenamefont {Pacati},\ and\ \citenamefont
  {Radici}}]{book}%
  \BibitemOpen
  \bibfield  {author} {\bibinfo {author} {\bibfnamefont {S.}~\bibnamefont
  {Boffi}}, \bibinfo {author} {\bibfnamefont {C.}~\bibnamefont {Giusti}},
  \bibinfo {author} {\bibfnamefont {F.~D.}\ \bibnamefont {Pacati}}, \ and\
  \bibinfo {author} {\bibfnamefont {M.}~\bibnamefont {Radici}},\ }\href@noop {}
  {\emph {\bibinfo {title} {Electromagnetic Response of Atomic Nuclei}}},\
  \bibinfo {series} {Oxford Studies in Nuclear Physics}, Vol.~\bibinfo {volume}
  {20}\ (\bibinfo  {publisher} {Clarendon Press},\ \bibinfo {address}
  {Oxford},\ \bibinfo {year} {1996})\BibitemShut {NoStop}%
\bibitem [{\citenamefont {Meucci}\ \emph {et~al.}(2009)\citenamefont {Meucci},
  \citenamefont {Caballero}, \citenamefont {Giusti}, \citenamefont {Pacati},\
  and\ \citenamefont {Ud\'{\i}as}}]{Meucci:2009nm}%
  \BibitemOpen
  \bibfield  {author} {\bibinfo {author} {\bibfnamefont {A.}~\bibnamefont
  {Meucci}}, \bibinfo {author} {\bibfnamefont {J.~A.}\ \bibnamefont
  {Caballero}}, \bibinfo {author} {\bibfnamefont {C.}~\bibnamefont {Giusti}},
  \bibinfo {author} {\bibfnamefont {F.~D.}\ \bibnamefont {Pacati}}, \ and\
  \bibinfo {author} {\bibfnamefont {J.~M.}\ \bibnamefont {Ud\'{\i}as}},\ }\href
  {\doibase 10.1103/PhysRevC.80.024605} {\bibfield  {journal} {\bibinfo
  {journal} {Phys. Rev. C}\ }\textbf {\bibinfo {volume} {80}},\ \bibinfo
  {pages} {024605} (\bibinfo {year} {2009})}\BibitemShut {NoStop}%
\bibitem [{\citenamefont {Meucci}\ \emph
  {et~al.}(2013{\natexlab{a}})\citenamefont {Meucci}, \citenamefont {Vorabbi},
  \citenamefont {Giusti}, \citenamefont {Pacati},\ and\ \citenamefont
  {Finelli}}]{esotici2}%
  \BibitemOpen
  \bibfield  {author} {\bibinfo {author} {\bibfnamefont {A.}~\bibnamefont
  {Meucci}}, \bibinfo {author} {\bibfnamefont {M.}~\bibnamefont {Vorabbi}},
  \bibinfo {author} {\bibfnamefont {C.}~\bibnamefont {Giusti}}, \bibinfo
  {author} {\bibfnamefont {F.~D.}\ \bibnamefont {Pacati}}, \ and\ \bibinfo
  {author} {\bibfnamefont {P.}~\bibnamefont {Finelli}},\ }\href {\doibase
  10.1103/PhysRevC.87.054620} {\bibfield  {journal} {\bibinfo  {journal} {Phys.
  Rev. C}\ }\textbf {\bibinfo {volume} {87}},\ \bibinfo {pages} {054620}
  (\bibinfo {year} {2013}{\natexlab{a}})}\BibitemShut {NoStop}%
\bibitem [{\citenamefont {Meucci}\ \emph {et~al.}(2004)\citenamefont {Meucci},
  \citenamefont {Giusti},\ and\ \citenamefont {Pacati}}]{Meucci:2003cv}%
  \BibitemOpen
  \bibfield  {author} {\bibinfo {author} {\bibfnamefont {A.}~\bibnamefont
  {Meucci}}, \bibinfo {author} {\bibfnamefont {C.}~\bibnamefont {Giusti}}, \
  and\ \bibinfo {author} {\bibfnamefont {F.~D.}\ \bibnamefont {Pacati}},\
  }\href {\doibase 10.1016/j.nuclphysa.2004.04.108} {\bibfield  {journal}
  {\bibinfo  {journal} {Nuclear Physics A}\ }\textbf {\bibinfo {volume}
  {739}},\ \bibinfo {pages} {277} (\bibinfo {year} {2004})}\BibitemShut
  {NoStop}%
\bibitem [{\citenamefont {Meucci}\ \emph
  {et~al.}(2011{\natexlab{b}})\citenamefont {Meucci}, \citenamefont
  {Caballero}, \citenamefont {Giusti},\ and\ \citenamefont
  {Ud\'{\i}as}}]{Meucci:2011pi}%
  \BibitemOpen
  \bibfield  {author} {\bibinfo {author} {\bibfnamefont {A.}~\bibnamefont
  {Meucci}}, \bibinfo {author} {\bibfnamefont {J.~A.}\ \bibnamefont
  {Caballero}}, \bibinfo {author} {\bibfnamefont {C.}~\bibnamefont {Giusti}}, \
  and\ \bibinfo {author} {\bibfnamefont {J.~M.}\ \bibnamefont {Ud\'{\i}as}},\
  }\href {\doibase 10.1103/PhysRevC.83.064614} {\bibfield  {journal} {\bibinfo
  {journal} {Phys. Rev. C}\ }\textbf {\bibinfo {volume} {83}},\ \bibinfo
  {pages} {064614} (\bibinfo {year} {2011}{\natexlab{b}})}\BibitemShut
  {NoStop}%
\bibitem [{\citenamefont {Meucci}\ and\ \citenamefont
  {Giusti}(2012)}]{Meucci:ant}%
  \BibitemOpen
  \bibfield  {author} {\bibinfo {author} {\bibfnamefont {A.}~\bibnamefont
  {Meucci}}\ and\ \bibinfo {author} {\bibfnamefont {C.}~\bibnamefont
  {Giusti}},\ }\href {\doibase 10.1103/PhysRevD.85.093002} {\bibfield
  {journal} {\bibinfo  {journal} {Phys. Rev. D}\ }\textbf {\bibinfo {volume}
  {85}},\ \bibinfo {pages} {093002} (\bibinfo {year} {2012})}\BibitemShut
  {NoStop}%
\bibitem [{\citenamefont {Meucci}\ \emph
  {et~al.}(2013{\natexlab{b}})\citenamefont {Meucci}, \citenamefont {Giusti},\
  and\ \citenamefont {Vorabbi}}]{Meucci:2013gja}%
  \BibitemOpen
  \bibfield  {author} {\bibinfo {author} {\bibfnamefont {A.}~\bibnamefont
  {Meucci}}, \bibinfo {author} {\bibfnamefont {C.}~\bibnamefont {Giusti}}, \
  and\ \bibinfo {author} {\bibfnamefont {M.}~\bibnamefont {Vorabbi}},\ }\href
  {\doibase 10.1103/PhysRevD.88.013006} {\bibfield  {journal} {\bibinfo
  {journal} {Phys. Rev. D}\ }\textbf {\bibinfo {volume} {88}},\ \bibinfo
  {pages} {013006} (\bibinfo {year} {2013}{\natexlab{b}})}\BibitemShut
  {NoStop}%
\bibitem [{\citenamefont {Meucci}\ and\ \citenamefont
  {Giusti}(2014{\natexlab{a}})}]{PhysRevD.89.117301}%
  \BibitemOpen
  \bibfield  {author} {\bibinfo {author} {\bibfnamefont {A.}~\bibnamefont
  {Meucci}}\ and\ \bibinfo {author} {\bibfnamefont {C.}~\bibnamefont
  {Giusti}},\ }\href {\doibase 10.1103/PhysRevD.89.117301} {\bibfield
  {journal} {\bibinfo  {journal} {Phys. Rev. D}\ }\textbf {\bibinfo {volume}
  {89}},\ \bibinfo {pages} {117301} (\bibinfo {year}
  {2014}{\natexlab{a}})}\BibitemShut {NoStop}%
\bibitem [{\citenamefont {Meucci}\ \emph
  {et~al.}(2011{\natexlab{c}})\citenamefont {Meucci}, \citenamefont {Giusti},\
  and\ \citenamefont {Pacati}}]{Meucci:2011nc}%
  \BibitemOpen
  \bibfield  {author} {\bibinfo {author} {\bibfnamefont {A.}~\bibnamefont
  {Meucci}}, \bibinfo {author} {\bibfnamefont {C.}~\bibnamefont {Giusti}}, \
  and\ \bibinfo {author} {\bibfnamefont {F.~D.}\ \bibnamefont {Pacati}},\
  }\href {\doibase 10.1103/PhysRevD.84.113003} {\bibfield  {journal} {\bibinfo
  {journal} {Phys. Rev. D}\ }\textbf {\bibinfo {volume} {84}},\ \bibinfo
  {pages} {113003} (\bibinfo {year} {2011}{\natexlab{c}})}\BibitemShut
  {NoStop}%
\bibitem [{\citenamefont {Gonz\'alez-Jim\'enez}\ \emph
  {et~al.}(2013)\citenamefont {Gonz\'alez-Jim\'enez}, \citenamefont
  {Caballero}, \citenamefont {Meucci}, \citenamefont {Giusti}, \citenamefont
  {Barbaro}, \citenamefont {Ivanov},\ and\ \citenamefont
  {Ud\'{\i}as}}]{PhysRevC.88.025502}%
  \BibitemOpen
  \bibfield  {author} {\bibinfo {author} {\bibfnamefont {R.}~\bibnamefont
  {Gonz\'alez-Jim\'enez}}, \bibinfo {author} {\bibfnamefont {J.~A.}\
  \bibnamefont {Caballero}}, \bibinfo {author} {\bibfnamefont {A.}~\bibnamefont
  {Meucci}}, \bibinfo {author} {\bibfnamefont {C.}~\bibnamefont {Giusti}},
  \bibinfo {author} {\bibfnamefont {M.~B.}\ \bibnamefont {Barbaro}}, \bibinfo
  {author} {\bibfnamefont {M.~V.}\ \bibnamefont {Ivanov}}, \ and\ \bibinfo
  {author} {\bibfnamefont {J.~M.}\ \bibnamefont {Ud\'{\i}as}},\ }\href
  {\doibase 10.1103/PhysRevC.88.025502} {\bibfield  {journal} {\bibinfo
  {journal} {Phys. Rev. C}\ }\textbf {\bibinfo {volume} {88}},\ \bibinfo
  {pages} {025502} (\bibinfo {year} {2013})}\BibitemShut {NoStop}%
\bibitem [{\citenamefont {Meucci}\ and\ \citenamefont
  {Giusti}(2014{\natexlab{b}})}]{PhysRevD.89.057302}%
  \BibitemOpen
  \bibfield  {author} {\bibinfo {author} {\bibfnamefont {A.}~\bibnamefont
  {Meucci}}\ and\ \bibinfo {author} {\bibfnamefont {C.}~\bibnamefont
  {Giusti}},\ }\href {\doibase 10.1103/PhysRevD.89.057302} {\bibfield
  {journal} {\bibinfo  {journal} {Phys. Rev. D}\ }\textbf {\bibinfo {volume}
  {89}},\ \bibinfo {pages} {057302} (\bibinfo {year}
  {2014}{\natexlab{b}})}\BibitemShut {NoStop}%
\bibitem [{\citenamefont {Caballero}\ \emph {et~al.}(2005)\citenamefont
  {Caballero}, \citenamefont {Amaro}, \citenamefont {Barbaro}, \citenamefont
  {Donnelly}, \citenamefont {Maieron},\ and\ \citenamefont
  {Ud\'{\i}as}}]{Caballero:2005sn}%
  \BibitemOpen
  \bibfield  {author} {\bibinfo {author} {\bibfnamefont {J.~A.}\ \bibnamefont
  {Caballero}}, \bibinfo {author} {\bibfnamefont {J.~E.}\ \bibnamefont
  {Amaro}}, \bibinfo {author} {\bibfnamefont {M.~B.}\ \bibnamefont {Barbaro}},
  \bibinfo {author} {\bibfnamefont {T.~W.}\ \bibnamefont {Donnelly}}, \bibinfo
  {author} {\bibfnamefont {C.}~\bibnamefont {Maieron}}, \ and\ \bibinfo
  {author} {\bibfnamefont {J.~M.}\ \bibnamefont {Ud\'{\i}as}},\ }\href
  {\doibase 10.1103/PhysRevLett.95.252502} {\bibfield  {journal} {\bibinfo
  {journal} {Phys. Rev. Lett.}\ }\textbf {\bibinfo {volume} {95}},\ \bibinfo
  {pages} {252502} (\bibinfo {year} {2005})}\BibitemShut {NoStop}%
\bibitem [{\citenamefont {Megias}\ \emph
  {et~al.}(2014{\natexlab{b}})\citenamefont {Megias}, \citenamefont {Ivanov},
  \citenamefont {Gonz\'alez-Jim\'enez}, \citenamefont {Barbaro}, \citenamefont
  {Caballero}, \citenamefont {Donnelly},\ and\ \citenamefont
  {Ud\'{\i}as}}]{minerva-juan}%
  \BibitemOpen
  \bibfield  {author} {\bibinfo {author} {\bibfnamefont {G.~D.}\ \bibnamefont
  {Megias}}, \bibinfo {author} {\bibfnamefont {M.~V.}\ \bibnamefont {Ivanov}},
  \bibinfo {author} {\bibfnamefont {R.}~\bibnamefont {Gonz\'alez-Jim\'enez}},
  \bibinfo {author} {\bibfnamefont {M.~B.}\ \bibnamefont {Barbaro}}, \bibinfo
  {author} {\bibfnamefont {J.~A.}\ \bibnamefont {Caballero}}, \bibinfo {author}
  {\bibfnamefont {T.~W.}\ \bibnamefont {Donnelly}}, \ and\ \bibinfo {author}
  {\bibfnamefont {J.~M.}\ \bibnamefont {Ud\'{\i}as}},\ }\href {\doibase
  10.1103/PhysRevD.89.093002} {\bibfield  {journal} {\bibinfo  {journal} {Phys.
  Rev.}\ }\textbf {\bibinfo {volume} {D89}},\ \bibinfo {pages} {093002}
  (\bibinfo {year} {2014}{\natexlab{b}})}\BibitemShut {NoStop}%
\bibitem [{\citenamefont {Abe}\ \emph {et~al.}(2013{\natexlab{a}})\citenamefont
  {Abe} \emph {et~al.}}]{PhysRevD.87.092003}%
  \BibitemOpen
  \bibfield  {author} {\bibinfo {author} {\bibfnamefont {K.}~\bibnamefont
  {Abe}} \emph {et~al.} (\bibinfo {collaboration} {T2K Collaboration}),\ }\href
  {\doibase 10.1103/PhysRevD.87.092003} {\bibfield  {journal} {\bibinfo
  {journal} {Phys. Rev. D}\ }\textbf {\bibinfo {volume} {87}},\ \bibinfo
  {pages} {092003} (\bibinfo {year} {2013}{\natexlab{a}})}\BibitemShut
  {NoStop}%
\bibitem [{\citenamefont {Abe}\ \emph {et~al.}(2014{\natexlab{a}})\citenamefont
  {Abe} \emph {et~al.}}]{Abe:2014iza}%
  \BibitemOpen
  \bibfield  {author} {\bibinfo {author} {\bibfnamefont {K.}~\bibnamefont
  {Abe}} \emph {et~al.} (\bibinfo {collaboration} {T2K Collaboration}),\
  }\href@noop {} {\  (\bibinfo {year} {2014}{\natexlab{a}})},\ \Eprint
  {http://arxiv.org/abs/1411.6264} {arXiv:1411.6264 [hep-ex]} \BibitemShut
  {NoStop}%
\bibitem [{\citenamefont {Abe}\ \emph {et~al.}()\citenamefont {Abe} \emph
  {et~al.}}]{t2k-nuedata}%
  \BibitemOpen
  \bibfield  {author} {\bibinfo {author} {\bibfnamefont {K.}~\bibnamefont
  {Abe}} \emph {et~al.} (\bibinfo {collaboration} {T2K Collaboration}),\
  }\href@noop {} {}\bibinfo {howpublished} {Data release for T2K 2014 $\nu_e$
  CC inclusive cross-section measurement,
  \url{http://t2k-experiment.org/results/nd280-nue-xs-2014}}\BibitemShut
  {NoStop}%
\bibitem [{\citenamefont {Abe}\ \emph {et~al.}(2014{\natexlab{b}})\citenamefont
  {Abe} \emph {et~al.}}]{Abe:2014agb}%
  \BibitemOpen
  \bibfield  {author} {\bibinfo {author} {\bibfnamefont {K.}~\bibnamefont
  {Abe}} \emph {et~al.} (\bibinfo {collaboration} {T2K Collaboration}),\ }\href
  {\doibase 10.1103/PhysRevLett.113.241803} {\bibfield  {journal} {\bibinfo
  {journal} {Phys. Rev. Lett.}\ }\textbf {\bibinfo {volume} {113}},\ \bibinfo
  {pages} {241803} (\bibinfo {year} {2014}{\natexlab{b}})}\BibitemShut
  {NoStop}%
\bibitem [{\citenamefont {Serot}\ and\ \citenamefont
  {Walecka}(1986)}]{Serot:1984ey}%
  \BibitemOpen
  \bibfield  {author} {\bibinfo {author} {\bibfnamefont {B.~D.}\ \bibnamefont
  {Serot}}\ and\ \bibinfo {author} {\bibfnamefont {J.~D.}\ \bibnamefont
  {Walecka}},\ }\href@noop {} {\bibfield  {journal} {\bibinfo  {journal} {Adv.
  Nucl. Phys.}\ }\textbf {\bibinfo {volume} {16}},\ \bibinfo {pages} {1}
  (\bibinfo {year} {1986})}\BibitemShut {NoStop}%
\bibitem [{\citenamefont {Reinhard}(1989)}]{Rein:1989}%
  \BibitemOpen
  \bibfield  {author} {\bibinfo {author} {\bibfnamefont {P.~G.}\ \bibnamefont
  {Reinhard}},\ }\href@noop {} {\bibfield  {journal} {\bibinfo  {journal} {Rep.
  Prog. Phys.}\ }\textbf {\bibinfo {volume} {52}},\ \bibinfo {pages} {439}
  (\bibinfo {year} {1989})}\BibitemShut {NoStop}%
\bibitem [{\citenamefont {Ring}(1996)}]{Rign:1996}%
  \BibitemOpen
  \bibfield  {author} {\bibinfo {author} {\bibfnamefont {P.}~\bibnamefont
  {Ring}},\ }\href@noop {} {\bibfield  {journal} {\bibinfo  {journal} {Prog.
  Part. Nucl. Phys.}\ }\textbf {\bibinfo {volume} {37}},\ \bibinfo {pages}
  {193} (\bibinfo {year} {1996})}\BibitemShut {NoStop}%
\bibitem [{\citenamefont {Lalazissis}\ \emph {et~al.}(1997)\citenamefont
  {Lalazissis}, \citenamefont {K\"onig},\ and\ \citenamefont
  {Ring}}]{Lalazissis:1996rd}%
  \BibitemOpen
  \bibfield  {author} {\bibinfo {author} {\bibfnamefont {G.~A.}\ \bibnamefont
  {Lalazissis}}, \bibinfo {author} {\bibfnamefont {J.}~\bibnamefont {K\"onig}},
  \ and\ \bibinfo {author} {\bibfnamefont {P.}~\bibnamefont {Ring}},\ }\href
  {\doibase 10.1103/PhysRevC.55.540} {\bibfield  {journal} {\bibinfo  {journal}
  {Phys. Rev. C}\ }\textbf {\bibinfo {volume} {55}},\ \bibinfo {pages} {540 }
  (\bibinfo {year} {1997})}\BibitemShut {NoStop}%
\bibitem [{\citenamefont {Serot}\ and\ \citenamefont
  {Walecka}(1997)}]{Serot:1997xg}%
  \BibitemOpen
  \bibfield  {author} {\bibinfo {author} {\bibfnamefont {B.~D.}\ \bibnamefont
  {Serot}}\ and\ \bibinfo {author} {\bibfnamefont {J.~D.}\ \bibnamefont
  {Walecka}},\ }\href {\doibase 10.1142/S0218301397000299} {\bibfield
  {journal} {\bibinfo  {journal} {Int. J. Mod. Phys.}\ }\textbf {\bibinfo
  {volume} {E6}},\ \bibinfo {pages} {515} (\bibinfo {year} {1997})}\BibitemShut
  {NoStop}%
\bibitem [{\citenamefont {Cooper}\ \emph {et~al.}(1993)\citenamefont {Cooper},
  \citenamefont {Hama}, \citenamefont {Clark},\ and\ \citenamefont
  {Mercer}}]{Cooper:1993nx}%
  \BibitemOpen
  \bibfield  {author} {\bibinfo {author} {\bibfnamefont {E.~D.}\ \bibnamefont
  {Cooper}}, \bibinfo {author} {\bibfnamefont {S.}~\bibnamefont {Hama}},
  \bibinfo {author} {\bibfnamefont {B.~C.}\ \bibnamefont {Clark}}, \ and\
  \bibinfo {author} {\bibfnamefont {R.~L.}\ \bibnamefont {Mercer}},\ }\href
  {\doibase 10.1103/PhysRevC.47.297} {\bibfield  {journal} {\bibinfo  {journal}
  {Phys. Rev. C}\ }\textbf {\bibinfo {volume} {47}},\ \bibinfo {pages} {297}
  (\bibinfo {year} {1993})}\BibitemShut {NoStop}%
\bibitem [{\citenamefont {Cooper}\ \emph {et~al.}(2009)\citenamefont {Cooper},
  \citenamefont {Hama},\ and\ \citenamefont {Clark}}]{Cooper:2009}%
  \BibitemOpen
  \bibfield  {author} {\bibinfo {author} {\bibfnamefont {E.~D.}\ \bibnamefont
  {Cooper}}, \bibinfo {author} {\bibfnamefont {S.}~\bibnamefont {Hama}}, \ and\
  \bibinfo {author} {\bibfnamefont {B.~C.}\ \bibnamefont {Clark}},\ }\href
  {\doibase 10.1103/PhysRevC.80.034605} {\bibfield  {journal} {\bibinfo
  {journal} {Phys. Rev. C}\ }\textbf {\bibinfo {volume} {80}},\ \bibinfo
  {pages} {034605} (\bibinfo {year} {2009})}\BibitemShut {NoStop}%
\bibitem [{\citenamefont {Meucci}\ \emph {et~al.}(2014)\citenamefont {Meucci},
  \citenamefont {Vorabbi}, \citenamefont {Giusti}, \citenamefont {Pacati},\
  and\ \citenamefont {Finelli}}]{PhysRevC.89.034604}%
  \BibitemOpen
  \bibfield  {author} {\bibinfo {author} {\bibfnamefont {A.}~\bibnamefont
  {Meucci}}, \bibinfo {author} {\bibfnamefont {M.}~\bibnamefont {Vorabbi}},
  \bibinfo {author} {\bibfnamefont {C.}~\bibnamefont {Giusti}}, \bibinfo
  {author} {\bibfnamefont {F.~D.}\ \bibnamefont {Pacati}}, \ and\ \bibinfo
  {author} {\bibfnamefont {P.}~\bibnamefont {Finelli}},\ }\href {\doibase
  10.1103/PhysRevC.89.034604} {\bibfield  {journal} {\bibinfo  {journal} {Phys.
  Rev. C}\ }\textbf {\bibinfo {volume} {89}},\ \bibinfo {pages} {034604}
  (\bibinfo {year} {2014})}\BibitemShut {NoStop}%
\bibitem [{\citenamefont {Abe}\ \emph {et~al.}(2013{\natexlab{b}})\citenamefont
  {Abe} \emph {et~al.}}]{PhysRevD.87.012001}%
  \BibitemOpen
  \bibfield  {author} {\bibinfo {author} {\bibfnamefont {K.}~\bibnamefont
  {Abe}} \emph {et~al.} (\bibinfo {collaboration} {T2K Collaboration}),\ }\href
  {\doibase 10.1103/PhysRevD.87.012001} {\bibfield  {journal} {\bibinfo
  {journal} {Phys. Rev. D}\ }\textbf {\bibinfo {volume} {87}},\ \bibinfo
  {pages} {012001} (\bibinfo {year} {2013}{\natexlab{b}})}\BibitemShut
  {NoStop}%
\bibitem [{\citenamefont {Martini}\ and\ \citenamefont
  {Ericson}(2014)}]{Martini:2014dqa}%
  \BibitemOpen
  \bibfield  {author} {\bibinfo {author} {\bibfnamefont {M.}~\bibnamefont
  {Martini}}\ and\ \bibinfo {author} {\bibfnamefont {M.}~\bibnamefont
  {Ericson}},\ }\href {\doibase 10.1103/PhysRevC.90.025501} {\bibfield
  {journal} {\bibinfo  {journal} {Phys. Rev. C}\ }\textbf {\bibinfo {volume}
  {90}},\ \bibinfo {pages} {025501} (\bibinfo {year} {2014})}\BibitemShut
  {NoStop}%
\bibitem [{\citenamefont {Amaro}\ \emph
  {et~al.}(2011{\natexlab{b}})\citenamefont {Amaro}, \citenamefont {Barbaro},
  \citenamefont {Caballero}, \citenamefont {Donnelly},\ and\ \citenamefont
  {Williamson}}]{Amaro:2010sd}%
  \BibitemOpen
  \bibfield  {author} {\bibinfo {author} {\bibfnamefont {J.~E.}\ \bibnamefont
  {Amaro}}, \bibinfo {author} {\bibfnamefont {M.~B.}\ \bibnamefont {Barbaro}},
  \bibinfo {author} {\bibfnamefont {J.~A.}\ \bibnamefont {Caballero}}, \bibinfo
  {author} {\bibfnamefont {T.~W.}\ \bibnamefont {Donnelly}}, \ and\ \bibinfo
  {author} {\bibfnamefont {C.~F.}\ \bibnamefont {Williamson}},\ }\href
  {\doibase 10.1016/j.physletb.2010.12.007} {\bibfield  {journal} {\bibinfo
  {journal} {Phys. Lett. B}\ }\textbf {\bibinfo {volume} {696}},\ \bibinfo
  {pages} {151} (\bibinfo {year} {2011}{\natexlab{b}})}\BibitemShut {NoStop}%
\end{thebibliography}
%

\end{document}